\documentstyle[12pt,aaspp]{article}
\begin{document}
%
\title{On Scaling Laws and Alfv\'{e}nic Magnetic Fluctuations in Molecular Clouds}

\author{Taoling Xie}
\affil{Laboratory for Millimeter-wave Astronomy, Department of Astronomy, University of Maryland, College Park, MD 20742; email: tao@astro.umd.edu}
\begin{center}
{\bf To appear in Astrophysical Journal Letters}
\end{center}
\begin{abstract}
Under the basic assumption that the observed turbulent motions in molecular clouds are
Alfv\'{e}nic waves or turbulence, we emphasize that the Doppler broadening of 
molecular line profiles directly measures the velocity amplitudes of the waves 
instead of the Alfv\'{e}n velocity. Assuming an equipartition between the kinetic energy and the Alfv\'{e}nic magnetic energy, we further propose the hypothesis that observed standard scaling laws in molecular clouds 
imply a roughly scale-independent fluctuating magnetic field, which might be
understood as a result of strong wave-wave interactions and subsequent energy cascade.
We predict that $\sigma_{v}\propto \rho^{-0.5}$ is a more basic and robust relation
in that it may approximately hold in any regions where the spatial energy density 
distribution is primarily determined by wave-wave interactions, including gravitationally unbound 
regions. We also discuss the fact that a scale-independent $\sigma_{B}^{2}$ 
appears to contradict existing 1-D and 2-D computer simulations of MHD turbulence in molecular clouds.
\end{abstract}
\section{INTRODUCTION}
Motions of material within interstellar molecular clouds have long been known 
to be dominantly turbulent except in small low-mass dense cores. 
A significant advance in the study of turbulence and cloud support was the 
finding of a set of scaling relations, i.e., correlations among cloud 
quantities such as the characteristic radius $R$, mean non-thermal velocity
dispersion $\sigma_{v}$ and the average density $n_{H_{2}}=\rho/m$,
for various samples of molecular 
structures ranging from $0.2\; pc$ to $200 \; pc$ in size (Larson 1981; 
Leung et al 1982; Myers 1983; Dame et al 1986; Scoville \& Sanders 1987; Solomon et al 1987; Falgarone, Puget \& Perault 1992; Blitz 1993), 
\begin{equation}
\sigma_{v} \propto R^{\alpha}, \alpha=0.5,
\end{equation}
\begin{equation}
n_{H_{2}} \propto R^{-\gamma}, \gamma=1,
\end{equation}
and
\begin{equation}
\sigma_{v} \propto n_{H_{2}}^{-\beta}, \beta=0.5.
\end{equation}
Only two of the above relations are independent; the power-law indexes 
are related by $\alpha=\beta\gamma$. Since the above values for $\alpha$, $\beta$ and $\gamma$ are controversial, we will refer Equations (1)-(3) as 
the Standard Scaling Laws. Largely established on a cloud-cloud basis 
(i.e., based on samples of clouds), the physical 
significance of these scaling laws lies in the natural extrapolation that 
they are on average a manifestation of self-similar individual clouds satisfying
these relations (Larson 1981), although arguably these correlations can be 
due to observational selection effects or even artifacts (Kegel 1989; Scalo 1990; 
Myers 1991; V\'{a}zquez-Semadeni et al 1996). In general, observational determination of these
scaling relations is based on average properties of clouds and subject to numerous
uncertainties. Therefore these scaling laws are necessarily rough and perhaps only of
value dimensionally in a strict sense. On the other hand, the large number of
observational studies that do find more or less the same scaling relations indicate
that these scaling laws may indeed tell us something physical (Scalo 1987; McKee 1989), at least in a 
dimensional sense. 

Proposed explanations to the scaling laws bifurcate into hydrodynamic and 
magnetohydrodynamic (MHD) regimes and were critically reviewed by Scalo (1987). Most of them converge on that one of the two independent 
scaling relations is due to the trend for virial equilibrium.
The hydrodynamic approach takes the scaling laws as
an evidence for energy cascade in strong eddy turbulence over a large range of scales,
similar to the Kolmogorov law for incompressible turbulence (Larson 1981), which
gives dimensionally $\sigma_{v}\sim R^{1/3}$. In the MHD regime, however, comparably 
rigorous theoretical approach has been hampered by the lack of theories capable of 
describing strong compressible MHD turbulence. 

\section{EVIDENCE FOR A SCALE-INDEPENDENT FLUCTUATING MAGNETIC FIELD}
Given a relatively strong interstellar magnetic field
and a moderate ionization fraction of $10^{-7}$ to $10^{-4}$ maintained by 
penetrating cosmic rays and UV photons, the gas in molecular clouds is expected 
to behave much like plasma which is subject to numerous instabilities.
If one recognizes the existence of a general interstellar magnetic field
and turbulence, then it seems difficult to reject the likelihood that the turbulent 
motions (even if it is initially purely hydrodynamic) further generate MHD waves such as
Alfv\'{e}n waves. In fact, Alfv\'{e}n waves have long been proposed as a viable
physical process for explaining the observed ``turbulence" in molecular clouds
(Arons and Max 1975, hereafter AM75; Zweibel \& Josafatsson 1983). Identifying (implicitly or explicitly) the turbulent velocity as 
the Alfv\'{e}n velocity, it has been further proposed that the Standard Scaling Laws can 
be explained if kinetic energy, energy of the general magnetic field and gravitational energy
are all in approximate equipartition and that magnetic field field $B$ is largely 
a constant over the scales concerned (Pellegatti Franco et al 1985; Scalo 1987; Myers \& Goodman 1988, hereafter MG88; Fleck 1988). 
The difficulty of understanding such a constant $B$, however, has been noted (Scalo 1987; MG88; Mouschovias \& 
Psaltis 1995); it contradicts not only the ``frozen-in" picture for the 
interstellar magnetic field (cf.\ Mouschovias 1987) but also the observational evidence for a $B-n$
scaling relationship (Troland \& Heiles 1986; Heiles et al 1993).

The Doppler broadening of the spectral line profiles, nevertheless, reflects the 
velocity amplitudes of the waves (i.e. bulk moving velocity of the particles), 
not the phase velocity $V_{A}$. The velocity amplitude of an Alfv\'{e}n wave packet of scale $\sim R$
is determined by the assumed equipartition between the kinetic energy density and the 
fluctuating magnetic energy density over the scale $R$ (AM75)
\begin{equation}
\frac{1}{2}\rho\sigma_{v}^{2}=\frac{\sigma_{B}^{2}}{8\pi}\; or\; \delta{v} \simeq \frac{\delta B}{(4\pi\rho)^{1/2}},
\end{equation}
where $\delta v=(8ln2)^{1/2}\sigma_{v}$ and $\delta B=(8ln2)^{1/2}\sigma_{B}$.\footnote{A Gaussian distribution 
is assumed for the fluctuating magnetic field and the velocity field with root-mean-squares $\sigma_{B}$ and $\sigma_{v}$. Note also that all the physical quantities in Equation (4) are implicitly associated with scale $R$.} Alfv\'{e}n (1953)
was perhaps the first to apply this relation astrophysically (solar granulation).
McKee \& Zweibel (1995) and Zweibel \& McKee (1995) have recently demonstrated this relationship for weak MHD 
turbulence (when $\delta B$ is much smaller than $B$). Although it is difficult to prove the validity of this relation 
in the strong compressible turbulence regime, 
both observational and theoretical studies of solar wind 
turbulence have confirmed this basic relation, which has become a basic building block of a lot of analyses (cf.\ Tu \& Marsch 1995; Lau \& Siregar 1996).
Recent MHD turbulence simulations for molecular clouds also verified
this energy equipartition (Passot et al 1995; Gammie \& Ostriker 1996). 

Next, let us look at the commonly assumed virial equilibrium for gravitationally bound clouds.  Observationally,
virial equilibrium is established by comparing the kinetic and the gravitational 
energy of gas (cf.\ Larson 1981; Blitz 1993). Theoretically, however, the situation is
much less trivial, as the presence of the general magnetic field should also be taken
into account (cf.\ Nakano 1984; Mouschovias 1987; McKee et al 1995). Naively, the observationally established virial 
equilibrium seems to underscore the direct role of the general magnetic field, which is
consistent with a possibility first raised by Mestel (1965) that the volume and
surface terms due to the general magnetic field have opposite signs and thus
tend to cancel. A potentially more interesting (and radical) possibility is that turbulent motions can
effectively reduce the effective pressure or stress of the magnetic field (Kleeorin et al 1996).
The gravitational energy density of a gas parcel at 
radius $R$ from the center of gravity can be written as $U_{G}=3CG\frac{M(R)\rho}{R}$,
where $C$ is a numerical factor depending mainly on the density distribution 
and $M(R)$ is the mass inside a radius $R$. The simplest form for the virial equilibrium in accord with
observations can be written as 
\begin{equation}
CG\frac{M\rho}{R}=\rho\sigma_{v}^{2}+\frac{\rho kT}{m},
\end{equation}
where $m$ is mean mass of the molecules.  This is the same approach as adopted by
MG88 and Caselli \& Myers (1995).
Defining a mean density $\rho_{m}$ so that $M(R)=\frac{4}{3}\pi \rho_{m} R^{3}$,
the left hand side of Equation (5) becomes $CG\rho_{m}\rho R^{2}$.
If $\rho_{m}\sim \rho$ within the errors of observational measurements, 
Equations (4) \& (5) lead to
\begin{equation}
 \rho=(\frac{4}{3}\pi CG)^{-1/2}(\frac{\sigma_{B}^{2}}{4\pi}+\frac{\rho}{m}kT)^{1/2}R^{-1}.
\end{equation}
Equations (4) and (6) give
\begin{equation}
\sigma_{v}=(\frac{4\pi}{3}CG)^{1/4}\frac{\frac{\sigma_{B}}{(4\pi)^{1/2}}}{(\frac{\sigma_{B}^{2}}{4\pi}+\frac{\rho}{m}kT)^{1/4}}R^{0.5}
\end{equation}  

It is clear that Equations (7),(6) and (4) match the Standard Scaling Laws,
Equations (1), (2) and (3), respectively, if and only if $\sigma_{B}^{2}$ is
scale-independent and thermal energy is small compared to the non-thermal energy.
Since the turbulent pressure in this case is $P_{turb}\simeq \rho \sigma_{v}^{2}=\frac{\sigma_{B}^{2}}{4\pi}$,
a scale-independent $\sigma_{B}^{2}$ is consistent with the 
consensus that the scaling laws imply an approximate dynamic 
pressure equilibrium in molecular clouds (cf.\ Chi\`{e}ze 1987; Fleck 1988; Maloney 1988; Elmegreen 1989; Blitz 1993).
For a sample of galactic molecular clouds associated with HII regions, Solomon et al (1987) 
found a velocity dispersion-size relationship $\sigma_{v}(km\;s^{-1})=0.7(\frac{R}{1pc})^{0.5}$.  
From Equation (7), we derive $\sigma_{B}=10.6/C^{1/2} \; \mu\; gauss=24 \; \mu\; gauss$ for $C\sim 0.2$ (uniform density).  
If these clouds are centrally condensed ($C$ is somewhat larger), the required $\sigma_{B}$ will be correspondingly smaller.
Note, however, when thermal kinetic energy density is not negligible on 
small scales, both the density-size and the velocity-size relations will be different from the standard forms.
When thermal kinetic energy density dominates, Equation (6) yields
the classic isothermal $r^{-2}$ density distribution; otherwise 
the power-law index 
$\gamma$ for the density-size relation ought to be between $1$ and $2$, as discussed by
MG88, McKee (1989), Fuller \& Myers (1992) and Caselli \& Myers (1995).

\section{WAVE-WAVE INTERACTIONS: AN ANALOGY TO KOLMOGOROV CASCADE}

Since observations indicate that the turbulent kinetic energy is often comparable to the energy of 
the general background magnetic field in some molecular clouds (e.g., MG88; Crutcher et al 1993; 1994),
MHD waves at different wavelengths are expected to have strong mutual interactions and 
thus become strong MHD turbulence (cf.\ Myers \& Khersonsky 1995).
Intuitively, non-linear wave-wave interactions and consequent energy cascade largely 
dominate the spatial energy density distribution (cf.\ Sagdeev \& Galeev 1969; 
Elmegreen 1990; Biskamp 1994; Goldreich \& Sridhar 1995), similar to the Kolmogorov 
energy cascade for hydrodynamic eddy turbulence. 
Unfortunately,
an eigenvalue description that is used for weak MHD turbulence breaks
down in this case, leaving essentially no other effective means so far 
for establishing a working theory which could tell us what the spatial energy density distribution should be for strong compressible MHD turbulence. The theoretical
difficulty is clearly witnessed by the recent controversy over the Iroshnikov-Kraichnan 
theory for incompressible MHD turbulence (e.g., Goldreich \& Sridhar 1995; Ng \& Bhattacharjee 1996), which is a much more simplified problem than strong compressible MHD 
turbulence in molecular clouds. 

In this case it seems useful to see what observations tell us.
Specifically, as argued in the previous section, the observed scaling laws in
molecular clouds indicate a scale-independent fluctuating magnetic energy density $\sigma_{B}^{2}$.
In terms of energy spectrum, $P_{k}$,
i.e., energy per unit volume per unit wavenumber as a 
function of wavenumber $k\sim 1/R$, a scale-independent $\sigma_{B}^{2}$ 
implies a power-law $P_{k} \propto k^{-1}$. What is intriguing is that this $k^{-1}$ spectrum corresponds to equipartition of mode amplitudes (Biskamp 1994) or Raleigh-Jeans energy equipartition (Sagdeev \& Galeev 1969) in the
sense that the wave packets of different scales obtain an equal share of the 
available energy within a unit volume of space. This implies further
that wave packets on larger scales would dominate the total amount of energy.

Now, is such a spatial energy density distribution physically 
plausible ? We consider two aspects of the problem. First, 
the energy flux of an Alfv\'{e}n wave packet on a scale $R$ can be written as 
$S=V_{A}\frac{\sigma_{B}^{2}}{4\pi}$ (cf.\ Lau \& Siregar 1996). It is evident
that an ensemble of interacting wave packets traveling in all directions would 
yield a net energy transport flux $\Delta S$ only in the direction of the negative 
spatial gradient for the magnetic pressure or energy density $\sigma_{B}^{2}/4\pi$, and thus the developing
tendency is to reduce the spatial gradients for $\sigma_{B}^{2}$. Second, a 
main characteristic of MHD turbulence is the existence of inverse cascade,
i.e., energy cascade from small to large scales with the tendency to form self-organized large-scale coherent structures (Biskamp 1994), in contrast to the 
direct energy cascade (from large to small scales) in pure hydrodynamic 
turbulence. Therefore it is not 
unlikely that the interacting wave packets of different scales achieve a net
tendency for an equal share of the available energy density due to the effective
energy 
cascade in both the direct and inverse directions while the system is fully turbulent (cf.\ Biskamp 1994).

One point that appears interesting to us is that if the kinetic energy does remain in equipartition with the fluctuating magnetic energy, then perhaps a hydrodynamic 
description and a MHD description could lead to comparable physical insights to the turbulence in molecular clouds, at least from an energy point of view. The MHD approach, however, does offer certain 
advantages. One of these is that MHD waves may effectively transport energy with a high
speed and perhaps also a low dissipation rate without 
necessary bulky flow of gas, and yet achieve more-or-less a uniform pressure environment.
The disadvantage of the MHD approach is, however, its mathematical complexity.

Given that a scale-independent fluctuating magnetic 
field is physically plausible and the simplicity of it being able to explain the observed
Standard Scaling Laws as discussed in previous section is remarkable, we feel obligated
to propose this possibility as an empirical hypothesis for further observational and 
theoretical investigations.

\section{DISCUSSION}
\subsection{$\sigma_{B}$ versus $B$}
A fluctuating magnetic field component is 
not only implied by the observed ``turbulence" or hydromagnetic
waves (Alfv\'{e}n 1953; AM75), but also expected to explain why 
molecular clouds do not free-fall to the center of mass along the generally-ordered
magnetic field lines (cf.\ Shu et al 1987). Shu (1991) even 
suggested the intriguing possibility that an anisotropic fluctuating magnetic field
might explain the
reported prolateness of low mass cores.
The fluctuating magnetic field $\sigma_{B}$ is, nevertheless, conceptually different from 
the general magnetic field $B$, even in the extreme case that $\sigma_B \sim B$.
This raises the issue as for if the Standard Scaling Laws require a 
constant general magnetic field $B$ over a large range of scales, as some have 
discussed (e.g., Pellegatti Franco et al 1985; Scalo 1987; MG88; Fleck 1988).
The answer to this question is negative based on the above discussions, but 
a constant $B$ would not contradict the scaling laws.  A key point
is that a constant $B$ does not directly explain the Standard Scaling Laws without 
invoking $\sigma_{B}$ first, because the
Alfv\'{e}n velocity is not the same as the observed fluctuating turbulent velocity
amplitudes. In other words, even if $B$ is a constant over a large range of scales,
a separate physical mechanism would still have to be sought to produce a scale-independent
$\sigma_{B}$.
In practice, however, it is likely that $\sigma_{B}$ becomes comparable to $B$ 
on large scales.
From the empirical $B-n$ relation $B\simeq 1.5(\frac{n_{H_{2}}}{1\;cm^{-3}})^{1/2}\; \mu\; gauss$ (Heiles et al 1993) and the estimated $\sigma_{B}\sim 24\;\mu\;gauss$ (see Section 2), it 
can be crudely estimated that $\sigma_{B}$ becomes comparable to $B$ when $n_{H_{2}}\simeq 260 \; cm^{-3}$, $\sigma_{v}\simeq 2.3\; km\; s^{-1}$ and $R\simeq 10\; pc$. Taking this at its face value, it seems that $\sigma_{B}$ could be well below $B$, or equivalently $\sigma_{v}$ considerably smaller than the Alfv\'{e}n velocity on scales smaller than a few parsecs.
\subsection{Predictions}
Generally speaking, our principal expectation is that regions where the spatial
distribution of the turbulent energy density is dominantly determined
by wave-wave interactions are more likely to demonstrate the Standard Scaling Laws,
 while regions where this is not
the case are unlikely to do so.
Perhaps this is why 
the recent observational results (Plume et al 1996) of massive star forming cores do not agree with the standard scaling laws, as localized energy sources 
might dominate the energy density distributions over the scales of massive star 
formation.
Further, the velocity-density relation (Equation (3)) among the Standard Scaling Laws 
should be the most robust physical 
relation in the sense that it may be expected for any region where magnetic fluctuations
become scale-independent. In contrast, the standard density-size and velocity-size relations
require not only scale-independent magnetic field fluctuations,
but also a second physical relationship such as the virial equilibrium\footnote{Heithausen (1996) has recently reported a remarkable correlation $n\propto R^{-0.8}$ for high latitude clouds, which made him question
if virial equilibrium is the true reason for a $n\propto R^{-1}$ relation. It is not clear yet, however, if his assuming equal distance for all the high latitude clouds would yield the correlations artificially.}.
This prediction seems supported by the observations of Falgarone, Puget \& Perault (1992) which find a remarkable correlation $\sigma_{v}\propto n_{H_{2}}^{-0.5}$ for a sample of gravitationally unbound molecular structures, although it is possible that jump shocks
may also give rise to this relation (J.Scalo, private communication).

\subsection{Comparison with Existing 1-D and 2-D MHD Simulations}
Our hypothesis for a scale-independent fluctuating magnetic field, however, 
appears to contradict 
the recent 1-D and 2-D computer simulations for strong compressible MHD turbulence in molecular clouds. These simulations report a fluctuating magnetic spectrum (energy per unit 
volume per unit wavenumber), $P_{k}\propto k^{-t}$, where $t\sim 2.0-2.3$ for 1-D (Gammie \& Ostriker 1996) and $t\sim 1.85$ for 2-D (Passot, V\'{a}zquez-Semadeni \& Pouquet 1995).
This magnetic spectrum implies a fluctuating magnetic field, $\sigma_{B} \propto R^{(t-1)/2}$ (John Scalo, private communication). In other words, these simulations show a steeper
k-dependence for the magnetic spectrum than our hypothesis. Both simulations find essentially the same velocity spectrum (energy per unit mass per unit wavenumber) $E_{k}\propto k^{-2}$, which is dimensionally consistent with the
the observed standard velocity-size relation $\delta v\sim R^{1/2}$. It appears difficult
for the simulations to explain the other observed Standard Scaling Laws. The difficulty of
explaining the standard size-density relation in the 2-D simulations of Passot et al (1995), for example, 
is recently made clear by V\'{a}zquez-Semadeni et al (1996). These authors were then led 
to suggest that the standard velocity-size relation is the basic relation while the
density-size relation may be an observational artifact, in clear contrast 
with our predictions in this paper. This raises the question as for if a simple dimensional 
analysis from a pure energy consideration as we have done in this paper is a valid approach for the MHD turbulence
and if the equipartition between kinetic energy and fluctuating magnetic energy is 
a good assumption.
On the other hand, 1-D and 2-D MHD turbulence simulations
in solar wind studies have been known to produce energy spectra too steep to compare
with observations, possibly due to the restriction in dimensionality, enforced 
periodicity and other unrealistic simplifications (Marsch 1991; Tu \& Marsch 1995).
So it appears that the standard scaling laws and subsequently our hypothesis for a 
scale-independent fluctuating magnetic field in molecular clouds remain a challenge to
MHD simulations, or vice versa. We share the same hope with the solar physicists that future 
3-D MHD simulations will shed light on this important issue in MHD turbulence.

In conclusion, we proposed the hypothesis that strong wave-wave interactions have led to
a scale-independent distribution of fluctuating magnetic energy density 
and consequently the observed scaling laws in molecular clouds. We suggest that further 
careful observational studies of interstellar atomic/molecular clouds not in virial 
equilibrium or not harboring active star forming activities may serve as valuable direct tests of the basic ideas presented in this 
article, especially in comparison to the contradicting predictions of recent 1-D and 2-D 
computer simulations.

The author gratefully acknowledges kindest encouragement from Professors Chuan-Sheng Liu, Paul Goldsmith and Stuart Vogel.
He is grateful to John Scalo for a critical review of the 
manuscript and numerous stimulating discussions. 
He further thanks Chuan-sheng Liu, John Wang, Frank Shu, Pedro Safier, Neal Evans, Paul Goldsmith,
Phil Myers, Alyssa Goodman, Bruce Elmegreen, Dick Crutcher, Arieh K\"{o}nigl, Richard Larson, A. Bhattacharjee, Eve Ostriker and Enrique V\'{a}zquez-Semadeni for useful discussions. 
This research is supported in part by the NSF grant AST9314847 to the
Laboratory for Millimeter-wave Astronomy at the University of Maryland.


\begin{references}
\reference Alfv\'{e}n, H. 1953, Cosmical Electrodynamics, (Oxford, The University Press), 146
\reference Arons, J. \& Max, C.E. 1975, ApJ, 196, L77 (AM75)
\reference Biskamp, D. 1994, Phys. Rev. E, 50, 2702
\reference Blitz, L. 1993, in Protostars \& Planets III, ed. E.H. Levy \& J.I. Lunine (Tucson: The University of Arizona Press), 125
\reference Caselli, P., \& Myers, P.C. 1995, ApJ, 446, 665
\reference Chi\`{e}ze, J.-P. 1987; AA, 171, 225
\reference Crutcher, R.M., Troland, T.H., Goodman, A.A., Heiles, C., Kaz\`{e}s, \& Myers, P.C. 1993, ApJ, 407, 175
\reference Crutcher, R.M., Mouschovias, T.Ch., Troland, T.H., \& Ciolek, G.E. 1994, ApJ, 427, 839
\reference Dame, T., Elmegreen, B., Cohen, R., \& Thaddeus, P. 1986, ApJ, 305, 892
\reference Elmegreen, B.G. 1989, ApJ, 338, 178
\reference Elmegreen, B.G. 1990, ApJ, 361, L77
\reference Falgarone, E., Puget, J.-L., \& P\'{e}rault, M. 1992, AA, 257, 715
\reference Fleck, R.C. 1988, ApJ, 328, 299
\reference Fuller, G.A., \& Myers, P.C. 1992, ApJ, 384, 523
\reference Gammie, C.F., \& Ostriker, E.C. 1996, ApJ, 466, 814
\reference Goldreich,P., \& Sridhar, S. 1995, ApJ, 438, 763
\reference Heiles, C., Goodman, A.A., McKee, C.F., \& Zweibel, E.G. 1993, in Protostars \& Planets III, ed. E.H. Levy \& J.I. Lunine (Tucson: University of Arizona Press), 279
\reference Heithausen, A. 1996, AA, 314, 251
\reference Kegel, W.H. 1989, AA, 225, 517
\reference Kleeorin, N., Mond, M., \& Rogachevskii, I. 1996, AA, 307, 293
\reference Larson, R.B. 1981, MNRAS, 194, 809
\reference Lau, Y.-T., \& Siregar, E. 1996, ApJ, 465, 451
\reference Leung, C.M., Kutner, M.L., \& Mead, K.N. 1982, ApJ, 262, 583
\reference Maloney,P. 1988, ApJ, 334, 761
\reference Marsch, E. 1991, in Physics of the Inner Heliosphere: Particles, Waves and Turbulence, ed. R.Schwenn \& E. Marsch (New York: Springer), 159
\reference McKee, C.F. 1989, ApJ, 345, 782
\reference McKee, C.F., \& Zweibel, E.G. 1995, ApJ, 440, 686
\reference McKee, C.F., Zweibel, E.G., Goodman, A.A. \& Heiles, C. 1993, in Protostars \& Planets III, ed. E.H. Levy \& J.I. Lunine (Tucson: University of Arizona Press), 327
\reference Mestel, L. 1965, QJRAS, 6, 265
\reference Mouschovias, T.Ch. 1987, in Physical Processes in Interstellar Clouds, ed. G.E.Morfill \& M. Scholer (Dordrecht:Reidel), 453
\reference Mouschovias, T.Ch., \& Psaltis, D. 1995, ApJ, 444, L105
\reference Myers, P.C. 1983, ApJ, 270, 105
\reference Myers,P.C. 1991, in ``Fragmentation of Molecular Clouds and Star Formation", ed. E.Falgarone \& G.Duvert (Dordrecht:Kluwer), 71
\reference Myers, P.C., \& Goodman, A.A. 1988, ApJ, 329, 392 (MG88)
\reference Myers, P.C., \& Khersonsky, V.K. 1995, ApJ, 442, 186
\reference Nakano, T. 1984, Fundam. Cosmic Phys. 9, 139
\reference Ng, C.S., \& Bhattacharjee, A. 1996, ApJ, 465, 845
\reference Passot, T., V\'{a}zquez-Semadeni, E., \& Pouquet, A. 1995, ApJ, 455, 536
\reference Pellegatti Franco, G.A., Tarsia, R.D., \& Quiroga, R.J. 1985, ApSS, 111, 343
\reference Sagdeev, R.Z., \& Galeev, A.A. 1969, Nonlinear Plasma Theory, (New York: W.A. Benjamin, Inc.)
\reference Scalo, J.M. 1987, in Interstellar Processes, ed. D.J. Hollenbach \& H.A. Thronson, Jr. (Dordrecht:Reidel), 349
\reference Scalo, J.M. 1990, in Physical Processes in Fragmentation and Star Formation, ed. R. Capuzzo-Dolcetta, C. Chiosi \& A. di Fazio (Dordrecht: Kluwer), 151
\reference Scoville, N.Z. \& Sanders, D.B. 1987, in Interstellar Processes, ed.
D.J.Hollenbach \& H.A. Thronson, Jr., (Dordrecht:Reidel), 21
\reference Shu, F.H. 1991, in The Physics of Star Formation and Early Stellar Evolution, eds. C.J.Lada \& N.D.Kylafis (Dordrecht:Kluwer), 365
\reference Shu, F.H., Adams, F.C., \& Lizano, S. 1987, ARAA, 25, 23 
\reference Solomon,P.M.,Rivolo,A.R.,Barrett,J.,\& Yahil, A. 1987, ApJ, 319, 730 
\reference Troland, T.H., \& Heiles, C. 1986, ApJ, 301, 339
\reference Tu, C.-Y., \& Marsch, E. 1995, Sp. Sci. Rev., 73, 1
\reference V\'{a}zquez-Semadeni, E., Ballesteros-Paredes, \& Rodr'{i}quez, L.F. 1996, ApJ, in press
\reference Zweibel, E.G., \& Josafatsson, K. 1983, ApJ, 270, 511
\reference Zweibel, E.G., \& McKee, C.F. 1995, ApJ, 439, 779
\end{references}
\end{document}